\documentstyle[preprint,aps,eqsecnum]{revtex}
\begin{document}
\tighten
\title{\bf {Hamiltonian Analysis of the Deep Inelastic Structure Function}}
\author{\bf A. Harindranath\thanks{
On leave of absence from Saha Institute of Nuclear Physics, Calcutta, India}}
\address{
International Institute of Theoretical and Applied Physics \\
Iowa State University \\
123 Office and Lab Link \\
Ames, IA 50011
}
\author{\bf Rajen Kundu}
\address{
Saha Institute of Nuclear Physics\\
Sector 1, Block AF, Bidhan Nagar \\
Calcutta 700064 India\\
}
\date{31 May 1996}

\maketitle

 
\begin{abstract}
We investigate the feasibility of analyzing deep inelastic structure
functions in Hamiltonian formalism by combining the light-front BJL limit of
high energy amplitudes and the Fock space (multi-parton) 
description of hadrons.
This study is motivated by  
some of the theoretical questions emerging from the ongoing
nonperturbative/perturbative studies in light-front QCD and also by current 
problems in the interface of perturbative/nonperturbative QCD.
In this preliminary study, we address the unpolarized structure function
$F_2$.
Our starting point 
is the expression for the quark structure function as the Fourier
transform of the expectation value of the {\it good component}
of a bilocal current in the target state.
By expanding the target state in a set of multi-particle Fock states, 
the structure function is expressed as the sum of squares of multi-parton
wavefunctions integrated over independent longitudinal and transverse parton
momenta. Utilizing the fact that the
multi-particle Fock states are connected with each other by the light-front
QCD Hamiltonian, we study questions of cancellation of collinear
singularities, factorization of mass singularities, and the 
logarithmic scaling violations in the
Hamiltonian picture. 
In this paper the essential
features of the formalism are illustrated utilizing the calculation of 
the structure function of a dressed quark and the evolution of the valence
part of the $q {\bar q}$ bound state structure function  
up to order $g^2$.

\end{abstract}
\section{Introduction}
One of the earliest motivations for studying field theories in the 
light-front formulation came from investigations 
in current algebra \cite{Alfaro}. 
Also of significant importance were the observation of Bjorken 
scaling \cite{bjo} in deep 
inelastic scattering and the development of Feynman's parton model
\cite{photon}. At a 
former level, starting from the algebra of currents in light- front theory 
and utilizing Bjorken-Johnson-Low (BJL) limit of high energy scattering one 
could $``$derive exact scaling" and arrive at an expression for the scaling 
function as the Fourier transform of a bilocal current matrix element 
\cite{jackiw}. 
The structure and consequences of light-front current algebra played a 
significant role in proposing Quantum Chromodynamics (QCD) as the underlying 
theory of strong interactions \cite{Fri 1}. However, many questions regarding 
renormalization aspects remained unanswered.

Following the discovery of asymptotic freedom in QCD, scaling and
scaling violations were studied \cite{PGW} utilizing operator 
product expansions near the light cone and the theory of composite operator
renormalization. In this language the emphasis is on manifest gauge and
Lorenz covariance and the intuitive parton picture remains hidden. On the
other hand, immediately after the discovery of scaling and prior to the
discovery of asymptotic freedom, 
the utility of the infinite
momentum frame picture/light-front language in providing simple, intuitive 
description of high
energy phenomena was recognized and exploited by many 
authors \cite{dly,bks,ks,lip}.  
Thus the intuitive approach (physical gauge, light cone variables, ...) to
study various issues in deep inelastic scattering have been vigorously
pursued \cite{sop,CFP,DDT,CS,AHM} along with the operator product
 expansion formalism.
In this approach the covariant Feynman Diagram (with four momentum
conservation at vertices and off-mass shell particles in intermediate
states) is still employed. In contrast is the approach \cite{KJK,BR1}
using old-fashioned perturbation theory (three
momentum conservation at the vertices, on-mass shell particles in
intermediate states), light front gauge, and light front variables. As has
been stressed many times by Brodsky, Lepage and collaborators \cite{BR2}, 
it is the latter approach which keeps close
contact with the intuitive parton model ideas within field theory.       

Old-fashioned perturbation theory in the infinite momentum frame was
utilized by Drell, Levy, and Yan \cite{dly} to study scaling of the deep
inelastic structure function in field theory in the context of pseudoscalar
Yukawa model. After the emergence of QCD as the underlying theory of strong
interactions, various aspects of structure function have been studied in 
the old-fashioned light-front
perturbation theory formalism. For example,
1) questions concerning the evolution of distributions with emphasis on
end-point ($ x \rightarrow 1$) behavior \cite{BR1} (higher twist effects), 
2) phenomenological aspects of
higher twist effects \cite{BBG}, 
3) gluon distribution function \cite{glue}, 4) photon
distribution function \cite{bur}, and 5) theoretically motivated parameterizations of
parton distributions \cite{BBS}.    

The main advantages of the Brodsky-Lepage framework may be summarized as
follows. The light-front perturbation theory approach is the one closest to
the intuitive parton picture, while maintaining the field theoretic 
aspects of the
problem in tact. The framework makes use of the properties of multi-parton
wavefunctions which span both the perturbative and non-perturbative sector
of the theory. The calculation of certain observables sometimes simplifies
drastically compared to the other approaches to the problem. 
The utility of light front dynamics for the interpretation and 
the calculation of polarized structure functions and higher twist structure
functions in general has been stressed also by Jaffe, Ji and collaborators
\cite{jaffe}.
The disadvantages of the approach are
mainly associated with the complexities of renormalization in the
Hamiltonian framework. (For early work on renormalization of QED in the
infinite momentum frame see, for example, Ref. \cite{brs}.)

At present, two major approaches aimed at the non-perturbative study of field 
theories in the light-front 
formalism \cite{glazek} are the Discretized Light Cone
Quantization (DLCQ) \cite{bp} and Light-Front Tamm-Dancoff (LFTD)
approximation\cite{lftd}.     
Recently there has also been a proposal to study QCD in
the light front Hamiltonian approach with particular emphasis on the aspects
of renormalization \cite{Wil1}. Two of the most interesting features of light
front dynamics in the Hamiltonian language are light front power counting
\cite{Wil1,hari1} and the special aspects of longitudinal dynamics.  

Perturbative and non-perturbative aspects of longitudinal dynamics are of
interest both from theoretical and phenomenological considerations. The
problem associated with exactly zero longitudinal momentum (the so-called
$``$zero-mode" problem) has attracted lot of attention recently. The
divergences associated with very small longitudinal momentum also play a
major role in the approach of Ref. \cite{Wil1} in the context of
incorporating non-perturbative effects in the effective light-front
Hamiltonian via perturbation theory framework. Study of deep inelastic
structure function in the language of multi-parton wavefunctions offers an
opportunity to study directly the question of 
cancellation/non-cancellation of small longitudinal momentum divergences in
physical observables.

At small values of the longitudinal momentum (or equivalently at very
small values of $x$), the long (longitudinal) distance aspects of the
theory are probed and one enters the overlapping region of perturbative and
non-perturbative dynamics.
The small $x$ behavior of structure functions has attracted lot of
attention recently in the light of new data from the HERA facility.
A recent work of interest is that of Mueller \cite{mue} on small $x$
behavior using multi-parton light-front wavefunctions.

In this work we study various aspects of the deep inelastic structure
function $F_2$ utilizing light-front symmetries and multi-parton wavefunctions.
As is well known, some aspects of
the problem are easily probed in coordinate space and some
aspects are easily probed in momentum space.  
A convenient starting point to discuss features of deep inelastic scattering
in the Hamiltonian framework is provided by the expression for the structure
function as the Fourier transform of a bilocal matrix element. Among the
many derivations of this result that exist in the literature, is the 
one utilizing equal-$x^+$ light-front current algebra and the 
BJL limit \cite{jackiw}. In this approach (which closely follows the
original derivation of Bjorken in the infinite momentum frame)
partons are not introduced from
the beginning and $``$perfect" scaling emerges (after canonical manipulations)
as a result of special properties of the
equal $x^+$ commutators of currents. One essential feature of the 
commutators is the non-local behavior in the longitudinal
direction which is a consequence of the symmetries on
the light-front. Furthermore,  the current commutators are found to
have the same {\it form} in free and interacting theories. Thus BJL limit
together with light-front
current algebra allows us to have a starting point (which is very close to
the physical answer) for the Hamiltonian analysis of the structure function..

To address the problem of scaling violations in the Hamiltonian picture
and the associated renormalization, it
is convenient to introduce the constituents of the hadron.  
As is well known, one can make the synthesis between the current algebra
picture and the parton picture by expanding the state in terms of
multi-parton (constituent) wavefunctions \cite{leut}.  
The partons that
appear in this formalism need not be collinear or massless but they are
always on mass shell since we are using the Hamiltonian dynamics. 
The $F_2$ structure function appears as the sum of squares of multi-parton
wavefunctions integrated over independent longitudinal and transverse
momenta. The behavior of multi-parton wavefunctions is determined  by the
light-front Hamiltonian. 
In this
picture scaling violations are intimately associated with the
renormalization aspects of multi-parton wavefunctions.

Apart from the issues of renormalization, there is a question that
confronts any approach that attempts the description of structure function
in terms of constituent wavefunctions. Is it possible, even in principle, to
describe the hadron structure function in terms of a finite number of
constituent wave functions, since $``$ the average number is infinite"
\cite{photon,bjo2} according to the standard folklore. What we have to worry
about are the gluon and sea distribution functions and their physical
interpretation. To address this question, we need to take into
consideration, their definition in terms of the multi-parton wavefunctions,
 various physical constraints from the normalization of the hadron state, 
longitudinal momentum sum rule, etc. The approach using parton wavefunctions
offers the possibility to address this question.

It is of interest also to see how various issues like
the cancellation of collinear singularities, 
factorization (separation of soft and hard physical processes), suppression
of coherent (interference) effects in the hard processes, scale
evolution of parton densities, etc in perturbative QCD emerge in
the present formalism.

The plan of this paper is as follows. In Sec. II we discuss the general
feature  of the structure function $F_2$. Structure function of a dressed
quark is discussed in Sec. III to illustrate the calculational framework.
In Section IV we discuss the structure function of a meson-like bound state. 
Logarithmic scaling violations in the bound state structure function are
discussed in Section V. Discussion and summary is presented in Sec. VI.
Finally the details of the normalization of the state and longitudinal
momentum fraction sum rules are discussed in appendices A and B
respectively.   
\section{Unpolarized structure function $F_2$}
We start from the expression for the $F_2$ structure function 
(more precisely, the quark momentum fraction distribution function):
\begin{eqnarray}
{F_2(x) \over x} = {i \over 2 \pi} \int d \eta e^{-i \eta x} {\bar
V}^{1}(\eta). \label{first} 
\end{eqnarray}
The generalized form factor ${\bar V}^{1}$ is defined by
\begin{eqnarray}
\langle P \mid {\bar {\cal J}}^{+}(y \mid 0) \mid P \rangle = P^+ {\bar
V}^{1}(\eta) , 
\end{eqnarray}
where 
\begin{eqnarray}
{\bar {\cal J}}^+(y \mid 0) = { 1 \over 2 i} \big[
{\bar \psi}(y) \gamma^+ \psi(0) - {\bar \psi}(0) \gamma^+ \psi(y) \big] 
. \label{second}
\end{eqnarray}
The bilocality is only in the longitudinal direction, {\it i.e., } 
$y^+=0, y^\perp =0$.
Note that $x = - {q^+ \over P^+}$ and $\eta = { 1 \over 2} P^+ y^-$.
The field operator $ \psi^+(y)$ is given by \cite{WH1}
\begin{eqnarray}
  \psi^+(y) = \left ( \begin{array}{c}
                      \xi(y) \\ 
                       0
                         \end{array} \right )  
\end{eqnarray}
where
\begin{eqnarray} 
\xi(y) =  \sum_{s_1} \chi_{s_1} \int  [dp_1] \big [ b(p_1, s_1) 
e^{-{i \over 2}p_1^+y^-}
+ d^\dagger(p_1,-s_1) e^{{i \over 2}{p_1^+ y^-}} \big],
\end{eqnarray}
with $ [dp_1]$ defined in appendix A.
In this framework, the operator has a simple structure and all the
complexities are buried in the state $\mid P \rangle$. The state $\mid P
\rangle$ can be expanded in terms of multi-parton Fock space amplitudes
which are related to each other through the relativistic light-front 
version of the Schroedinger equation 
\begin{eqnarray}
P^- \mid P \rangle = { M^2 + (P^\perp)^2 \over P^+} \mid P \rangle ,
\label{rse}
\end{eqnarray}
where $P^-$ is the light-front QCD Hamiltonian which can be written as 
\begin{eqnarray}
P^- = P_0^- + V
\end{eqnarray}
with $P_0^-$ is the free part and $V$ is the interaction part. The explicit
expressions for $P_o^-$ and $V$ are given in Ref. \cite{WH1}. 
We note that the main focus of the ongoing effort in light-front theory is
to solve this equation nonperturbatively. We can explore various features of
the structure function utilizing the light-front Hamiltonian $P^-$.

When we address the problem of calculating the 
structure function of a composite system like a baryon
or a meson, which of course are the physically interesting systems, we are
immediately bombarded with all the complexities in the real world. To break
up the problem in to simpler pieces, in the following section 
we first consider an artificial target,
namely, a dressed parton. Even though such an object does not appear in the
real world, this calculation will help us to understand some of the key issues
in a simpler setting and also to set the notation. For 
calculations of structure function of a quark done 
in the framework of Feynman perturbation theory
in planar gauge see Dokshitzer, Dyakonov, and Troyan \cite{DDT}.    
\section{Structure function of a dressed quark}
\subsection{Parton picture}
We take the state $\mid P \rangle$ to be a dressed quark and expand this
state in terms of bare states of quark, quark plus gluon, quark plus two
gluons, etc. The expansion takes the form
\begin{eqnarray}
\mid P \sigma \rangle = && \phi_1(P,\sigma) b^\dagger(P,\sigma) \mid 0
\rangle \nonumber \\
&& ~ + \sum_{\sigma_1,\lambda_2} \int \{dk_1\} \int \{dk_2\}  \nonumber \\
&& ~~ \phi_2(P,\sigma  \mid k_1, \sigma_1; k_2 , \lambda_2)
\sqrt{2 (2 \pi)^3 P^+} \delta^3(P-k_1-k_2) b^\dagger(k_1,\sigma_1)
a^\dagger(k_2,\lambda_2) \mid 0 \rangle \nonumber \\
&& ~ + {1 \over 2} \sum_{\sigma_1, \lambda_2, \lambda_3} 
\int \{dk_1\} \int \{dk_2\} \int\{[dk_3\}  \nonumber \\
&& ~~\phi_3(P,\sigma \mid k_1, \sigma_1; k_2, \lambda_2; k_3, \lambda_3)
\sqrt{2 (2 \pi)^3 P^+} \delta^3(P-k_1-k_2-k_3)
\nonumber \\
&&~~~~~ b^\dagger(k_1, \sigma1) a^\dagger(k_2, \lambda_2) a^\dagger(k_3, \lambda_3)
\mid 0 \rangle \nonumber  \\
&& ~ + ~~~~~~... .\label{state}
\end{eqnarray}

The factor of ${1 \over 2}$ in front of the expression in the third term
above is the symmetry factor for identical bosons.

The function $\phi(P,\sigma)$ is the probability amplitude 
to find a bare quark with momentum $P$ and helicity $\sigma$ 
in a dressed quark, the function $\phi(P,\sigma \mid
k_1 \sigma_1, k_2 \lambda_2)$ is the probability amplitude to find a bare
quark with momentum $k_1$ and helicity $\sigma_1$ and a bare gluon 
with momentum $k_2$ and helicity $\lambda_2$ in the dressed quark, etc.

We evaluate the expression in Eq.(\ref{first}) explicitly, noting that in the 
present case
the contribution from the second term in Eq.(\ref{second}) is zero.
Introduce the Jacobi momenta ($x_i, \kappa_i^\perp$) 
\begin{eqnarray}
k_i^+ = x_i P^+, \,  k_i^\perp = \kappa_i^\perp + x_i P^\perp 
\end{eqnarray} 
so that 
\begin{eqnarray} 
\sum_i x_i = 1, \sum_i \kappa_i^\perp =0.
\end{eqnarray}
Also introduce the amplitudes,
\begin{eqnarray}
\phi_1 = && \psi_1, \nonumber \\
\phi_2(k_i^+, k_i^\perp) = && {1 \over \sqrt{P^+}} \psi_2(x_i,
\kappa_i^\perp),
\nonumber \\
\phi_3(k_i^+, k_i^\perp) = && { 1 \over P^+} \psi_3(x_i, \kappa_i^\perp),
\end{eqnarray}
and so on.
We get 
\begin{eqnarray}
{F_2(x) \over x} = && \mid \psi_1 \mid^2 \delta (1-x) \nonumber \\
&& + \sum_{\sigma_1, \lambda_2} \int dx_2 \int d^2 \kappa_1^\perp
\int d^2 \kappa_2^\perp \delta (1-x-x_2) \delta^2(\kappa_1^\perp+
\kappa_2^\perp) \mid \psi_2^{\sigma_1, \lambda_2}(x,\kappa_1^\perp; x_2,
\kappa_2^\perp) \mid^2 \nonumber \\
&& + {1 \over 2} \sum_{\sigma_1,\lambda_2,\lambda_3} \int dx_2 \int dx_3
\int d^2 \kappa_1^\perp \int d^2 \kappa_2^\perp \int d^2 \kappa_3^\perp
\delta (1-x-x_2-x_3) \delta^2(\kappa_1^\perp+\kappa_2^\perp+ \kappa_3^\perp)
\nonumber \\ 
&&~~~~~~~~~~~~~~~~~\mid \psi_3^{\sigma_1\lambda_2\lambda_3}(x,\kappa_1^\perp;
x_2, \kappa_2^\perp;
x_3 ,\kappa_3^\perp)\mid^2 \, \, + \, \, ... \label{third}
\end{eqnarray}
This equation makes manifest the partonic interpretation of the
quark distribution function, namely, the quark distribution function of a
dressed quark is the
incoherent sum of probabilities to find a bare parton (quark) with longitudinal
momentum fraction $x$ in various multi-particle Fock states of the dressed 
quark. Since we have computed the distribution function in field theory,
there are also significant
differences from the traditional parton model\cite{photon}. Most important
difference is the fact that the partons in field theory have transverse
momenta ranging from zero to infinity. Whether the structure function scales
or not now depends on the ultraviolet behavior of the multi-parton
wavefunctions. By analyzing various interactions, one easily finds that in
superrenormalizable interactions, the transverse momentum integrals 
converge in the ultraviolet and the structure function scales, 
whereas in renormalizable
interactions, the transverse momentum integrals diverge in the ultraviolet
which in turn leads to scaling violations in the structure function. 

We should also note that even though (\ref{third}) looks like an incoherent
sum, interference effects are also present in this expression. We will
elaborate more on this in Sec.V.           
 
\subsection{Dressing with one gluon}
In this section we evaluate the structure function of a dressed quark to
order $\alpha_s$. Our starting point is the eigenvalue equation obeyed by
the dressed quark state, {\it i.e.,} Eq.(\ref{rse}).
We substitute the Fock space expansion (\ref{state}) in (\ref{rse}) and make
projection with a bare one quark - one gluon state $ b^\dagger(p_1,s_1)
a^\dagger(p_2,\rho_2) \mid 0 \rangle$ keeping only terms up to order $g$. 
We arrive at
\begin{eqnarray}
\big[ {M^2 + (P^\perp)^2 \over P^+} - {m^2 + (p_1^\perp)^2 \over p_1^+} 
- {(p_2^\perp)^2 \over p_2^+} \big] \phi_2(P, \sigma \mid p_1,s_1; p_2,\rho_2)
= \nonumber \\
~~~~ - { g \over \sqrt{2 (2 \pi)^3}} T^a {1 \over \sqrt{p_2^+}}
\chi^\dagger_{s_1} 
\big[ 2 {p_2^\perp \over p_2^+} - {\sigma^\perp.p_1^\perp - im \over p_1^+}
\sigma^\perp - \sigma^\perp {\sigma^\perp.P^\perp + im \over P^+} 
\big] \chi_\sigma . {(\epsilon^\perp_{\rho_2})}^* ~ \phi_1(P,\sigma) . 
\end{eqnarray}
We rewrite the above equation in terms of Jacobi momenta
($ p_i^+ = x_i P^+, \kappa^\perp_i =p^\perp_i + x_i P^\perp$) and the
wavefunctions $\psi_i$ which are functions of Jacobi momenta. Using the 
notation $x=x_1, \kappa_1 = \kappa$ and using the facts $x_1+x_2=1, 
\kappa_1+\kappa_2=0$, we have
\begin{eqnarray}
\big[ M^2 - {m^2 +(\kappa^\perp)^2 \over x} - {(\kappa^\perp)^2 \over 
1-x} \big] \psi_2^{s_1, \rho21}(x,\kappa^\perp; 1-x, - \kappa^\perp) 
=  \nonumber \\
~~~ - { g \over \sqrt{2 (2 \pi)^3}} T^a {1  \over \sqrt{1-x}}
\chi^\dagger_{s_1} \big[ -2 {\kappa^\perp \over 1-x} -
{\sigma^\perp.\kappa^\perp -im \over x} \sigma^\perp - \sigma^\perp i m
\big] \chi_\sigma .{(\epsilon^\perp_{\rho_2})}^* \psi_1
\end{eqnarray} 
Taking the bare and dressed quarks to be massless,
we arrive at
\begin{eqnarray} 
\sum_{\sigma_1,\rho_2} \int d^2 \kappa^\perp \mid
\psi_2^{\sigma_1,\rho_2}(x,\kappa^\perp, 1-x, -\kappa^\perp) \mid^2 = 
{g^2 \over (2 \pi)^3} C_f \mid \psi_1 \mid^2 { 1 + x^2 \over 1-x} 
\int d^2 \kappa^\perp { 1 \over (\kappa^\perp)^2}
\end{eqnarray}    
where $C_f = {N^2 -1 \over 2N}$.
The transverse momentum integral is divergent at both limits of integration. We regulate the
lower limit by $\mu$ and the upper limit by $Q$. Thus we have
\begin{eqnarray}
{F_2(x) \over x} = \mid \psi_1 \mid^2 \big [ \delta (1-x) \, 
+ \, {\alpha_s \over 2 \pi}C_f{1+x^2 \over 1-x} ln{Q^2 \over \mu^2} \big].
\end{eqnarray}
The normalization condition reads 
\begin{eqnarray}
\mid \psi_1 \mid^2 \big[ 1 + {\alpha_s \over 2 \pi} C_f
\int dx{ 1 + x^2 \over 1-x}
ln{Q^2 \over \mu^2} \big ] = 1.
\end{eqnarray}
Within the present approximation (valid only upto $\alpha_s$),
\begin{eqnarray}
\mid \psi_1 \mid^2 = 1 - {\alpha_s \over 2 \pi} C_f \int dx{ 1 + x^2 \over 1-x}
ln{Q^2 \over \mu^2} .
\end{eqnarray}
In the second term we recognize the familiar expression of wavefunction 
correction of the state
$n$ in old fashioned perturbation theory, namely,  $ \sum'_m {\mid 
\langle m \mid V \mid n \rangle \mid^2 \over (E_n -E_m)^2}$.

Thus to order $\alpha_s$,
\begin{eqnarray}
{F_2(x) \over x} = \delta(1-x) + {\alpha_s \over 2 \pi} C_f ln{Q^2 \over \mu^2}
\big[ {1+x^2 \over 1-x} - \delta(1-x) \int dy {1+y^2 \over 1-y} \big].
\label{onel}
\end{eqnarray}
Note that  (\ref{onel}) can also be written as
\begin{eqnarray}
{F_2(x) \over x} = \delta(1-x) + {\alpha_s \over 2 \pi} C_f ln{Q^2 \over \mu^2}
\big[ {1 +x^2 \over (1-x)_{+}} + {3 \over 2} \delta(1-x) \big],
\end{eqnarray}
which is a more familiar expression.
Note that by construction $\mid \psi_2(x, \kappa^\perp) \mid^2$ is a
probability density. However, this function is singular as $x \rightarrow 1$
(gluon longitudinal momentum fraction approaching zero). To get a finite
probability density we have to introduce a cutoff $\epsilon$ 
($x_{gluon} > \epsilon)$, for example. In a physical cross section, this
$\epsilon $ cannot appear and here we have an explicit example of this
cancellation. 
The probabilistic interpretation of the splitting function 
$ P_{qq} = {1 +x^2 \over 1-x}$ which arise from real gluon emission 
is obvious in our derivation. On the other hand, the function 
${\tilde P}_{qq} =  {1 +x^2 \over (1-x)_{+}} + {3 \over 2} \delta(1-x) $
doesn't have the probabilistic interpretation since it includes contribution
from virtual gluon emission. This is immediately transparent from the
relation
\begin{eqnarray}
\int dx {\tilde P}_{qq}(x) = 0.
\end{eqnarray}
 We also note that the divergence arising from small transverse momentum
(the familiar mass singularity) cannot be handled properly in the present 
calculation. This is to be contrasted
with the calculation of the meson structure function (see the following
section) where the mass singularities can be properly absorbed into the
nonperturbative part of the structure function.    

It is instructive to compare the above derivation with typical 
calculations that
exist in the
literature. Conventionally, one evaluates the imaginary part of the forward
virtual Compton amplitude order by order in perturbation theory. In the lowest
order, in addition to the real gluon emission, one also has various self-energy
diagrams and the dressing of the photon vertex by the gluon. The ultraviolet
part of the vertex and
parts of the self-energy contributions cancel each other as a result of the
QED Ward Identity $Z_1=Z_2$. Thus, for example, in planar gauge, Dokshitzer,
Dyakonov and Troyan\cite{DDT} 
find by explicit calculations that neither collinear nor infrared 
mass singularities affect the
$``$partonometric vertex" and that the radiative corrections to the photon vertex
appear to be effectively unity. By explicit calculations both in the $\phi^3$
theory and QCD, Collins, Soper, and Sterman \cite{CSS} find that non-ladder 
diagrams are either
higher twist or contribute only to the hard part.
In the calculation of Brodsky and Lepage
\cite{BR1}
using light-front time ordered perturbation theory in the gauge $A^+=0$,
the identity $Z_1=Z_2$ is used. In the present calculation, since we 
have started
from the BJL limit, dressing of the photon
vertex, gluon radiation from the final state, and parts of self-energy 
contributions (that cancel eventually)
does not appear at all which leads to simplification of the calculation. It
is of interest to see whether such simplifications persist to higher orders
in the present framework.  
\section{Structure Function of a meson-like bound state}
For simplicity we consider a meson-like bound state. 
We expand the state $\mid P \rangle $ for $ q {\bar q}$ bound state in terms
of the Fock components $q {\bar q}$, $q {\bar q}g$, ... as follows.
\begin{eqnarray}
\mid P \rangle = && \sum_{\sigma_1, \sigma_2} \int [dk_1] \int [dk_2]
\nonumber \\
&& \phi_2(P \mid k_1, \sigma_1; k_2, \sigma_2) \sqrt{2 ((2 \pi)^3 P^+}
\delta^3(P-k_1-k_2) b^\dagger(k_1, \sigma_1) d^\dagger(k_2,\sigma_2) \mid 0
\rangle \nonumber \\
&& + \sum_{\sigma_1,\sigma_2,\lambda_3} \int \{dk_1\} \int \{dk_2\} \int 
\{dk_3\}
\nonumber \\
&& \phi_3(P \mid k_1, \sigma_1; k_2, \sigma_2; k_3, \lambda_3)
\sqrt{2 (2 \pi)^3 P^+} \delta^3(P-k_1 -k_2 -k_3) 
\nonumber \\
&&~~~~~b^\dagger(k_1 ,\sigma_1)
d^\dagger (k_2, \sigma_2) a^\dagger(k_3, \lambda_3) \mid 0 \rangle \nonumber
\\
&& + \, \, \, ... \, \, \, . \label{meson1}
\end{eqnarray}
Here $\phi_2$ is the probability amplitude to find a quark and an antiquark
in the meson, $\phi_3$ is the probability amplitude to find a quark,
antiquark and a gluon in the meson etc. 

As in Sec. III we evaluate the expression in Eq.(\ref{first}) explicitly.
The contribution from the first term (from the quark),
in terms of the amplitudes 
\begin{eqnarray}
\phi_2(k_i^+, k_i^\perp) = && { 1 \over \sqrt{P^+}}
 \psi_2 (x_i, \kappa_i^\perp), 
\nonumber \\
\phi_3(k_i^+, k_i^\perp) = && { 1 \over P^+} \psi_3
(x_i, \kappa_i^\perp),
\end{eqnarray}
and so on, is
\begin{eqnarray}
{F_2(x)^{q} \over x} = && \sum_{\sigma_1,\sigma_2} \int dx_2 \int
d^2\kappa_1^\perp \int d^2 \kappa_2^\perp \delta (1 - x -x_2)
\delta^2(\kappa_1 +  \kappa_2) \mid \psi_2^{\sigma_1,
\sigma_2}(x,\kappa_1^\perp; x_2 \kappa_2^\perp) \mid^2 \nonumber \\
&& + \sum_{\sigma_1, \sigma_2, \lambda_3} \int dx_2 \int dx_3 \int d^2
\kappa_1^\perp \int d^2 \kappa_2^\perp \int d^2 \kappa_3^\perp \delta(1 -
x -x_2 -x_3) \delta^2(\kappa_1 + \kappa_2 + \kappa_3)
\nonumber \\
&& ~~~~~~~~~~~~~~~~~~~~ \mid \psi_3^{\sigma_1, \sigma_2, \lambda_3}(x,
\kappa_1^\perp; x_2, \kappa_2^\perp; x_3, \kappa_3^\perp) \mid^2 + ... ~~ .
\label{exact}
\end{eqnarray}
Again, the partonic interpretation of the $F_2$ structure function is
manifest in this expression.

Contributions to the structure function from the second term in 
Eq.(\ref{first}) is
\begin{eqnarray}
{F_2(x)^{\bar q} \over x} = && \sum_{\sigma_1,\sigma_2} \int dx_2 \int
d^2\kappa_1^\perp \int d^2 \kappa_2^\perp \delta (1 - x -x_2)
\delta^2(\kappa_1 +  \kappa_2) \mid \psi_2^{\sigma_1,
\sigma_2}(x_2,\kappa_2^\perp; x, \kappa_1^\perp) \mid^2 \nonumber \\
&& + \sum_{\sigma_1, \sigma_2, \lambda_3} \int dx_2 \int dx_3 \int d^2
\kappa_1^\perp \int d^2 \kappa_2^\perp \int d^2 \kappa_3^\perp \delta(1 -
x -x_2 -x_3) \delta^2(\kappa_1 + \kappa_2 + \kappa_3)
\nonumber \\
&& ~~~~~~~~~~~~~~~~~~~~ \mid \psi_3^{\sigma_1, \sigma_2, \lambda_3}(x_2,
\kappa_2^\perp; x, \kappa_1^\perp; x_3, \kappa_3^\perp) \mid^2 + ... ~~ .
\end{eqnarray}
The normalization condition given in Eq.(\ref{norm}) guarantees that
\begin{eqnarray}
\int dx \big[ {F_2^{q}(x) \over x} + {F_2^{\bar q} (x) \over x} \big ] = 2
\end{eqnarray}
which reflects the fact that there are two valence particles in the meson.
Since the bilocal current component ${\bar {\cal J}}^+$ involves only
fermions explicitly, we appear to have missed the contributions from the
gluon constituents altogether. Gluonic contribution to the structure
function $F_2$ is most easily calculated by studying the hadron
expectation value of the conserved longitudinal momentum operator $P^+$. The
details are given in appendix C.

From the normalization condition, it is clear that the valence distribution
receives contribution from the amplitudes  $\psi_2$, $\psi_3$, ...  at any
scale $\mu$. This has interesting phenomenological implications. In the
model for the meson with only a quark-antiquark pair of equal mass, 
the valence distribution function will peak at $x = {1 \over 2}$. If there
are more than just the two particles in the system, the resulting valence
distribution will naturally have an enhancement in the $ x < {1 \over 2}$
region and a depletion in the $ x > { 1 \over 2} $ region, as a simple
consequence of longitudinal momentum conservation. 

The equation (\ref{exact}) as it stands is useful only when the bound state
solution in QCD is known in terms of the multi-parton wavefunctions. The
wavefunctions, as they stand,
span both the perturbative and non-perturbative sectors of the theory.
Great progress in the understanding of QCD in the high energy sector is made
in the past by separating the soft (non-perturbative) and hard
(perturbative) regions of QCD via the machinery of factorization. It is of
interest to see under what circumstances 
a factorization occurs in the formal result of Eq.
(\ref{exact}) and a perturbative picture of scaling violations emerges
finally. We address this issue in the following section.

\section{Perturbative picture of scaling violations}
To address the issue of scaling violations in the structure function
of the "meson-like" bound state, it is convenient to separate the momentum
space into low-energy and high-energy sectors. Such a separation has been
introduced in the past in the study of renormalization aspects of bound
state equations \cite{yukawa} in light-front field theory. The two sectors
are formally defined by introducing cutoff factors in the momentum space
integrals. How to cutoff the momentum integrals in a sensible and convenient
way in light-front theory is a subject under active research at the 
present time. Complications arise because of the possibility of large energy
divergences from both small $k^+$ and large $k^\perp$ regions.
In the following we investigate only the effects of logarithmic
divergences arising from
large transverse momenta, ignore subtelities arising from both small $x$
($x \rightarrow 0$)
and large $x$ ($x \rightarrow 1$) regions and subsequently use simple
transverse momentum cutoff. For complications arising from $x
\rightarrow 1$ region see Ref. \cite{BR1}.      
\subsection{Scale separation}
We define the soft region to be $\kappa^\perp < \mu$ and the hard
region to be $ \mu < \kappa^\perp < \Lambda$. 
$\mu$ serves as a factorization scale which separates soft and hard regions.
Since it is an intermediate scale introduced artificially purely for
convenience, physical structure function should be independent of $\mu$.
The multi-parton amplitude $\psi_2$ is a function of a single relative
transverse momentum $\kappa_1^\perp$ and we define
\begin{eqnarray}
 \psi_2 = \left \{ \begin{array}{c} \psi_2^{s}, 
~ 0 < \kappa_1^\perp < \mu, \\
     \psi_2^{h}, ~~~ \mu < \kappa_1^\perp < \Lambda \end{array} \right.
\end{eqnarray}  
The amplitude $\psi_3$ is a function of two relative momenta,
$\kappa_1^\perp$ and $\kappa_2^\perp$ and  we define
\begin{eqnarray}
\psi_3 = \left \{ \begin{array} {c} 
\psi_3^{ss}, ~ 0 < \kappa_1^\perp, \kappa_2^\perp < \mu \\
\psi_3^{sh}, ~ 0 < \kappa_1^\perp < \mu, \\
\psi_3^{hs},~ \mu < \kappa_1^\perp < \Lambda, ~ 0 <
\kappa_2^\perp < \mu, \\
\psi_3^{hh}, ~ \mu < \kappa_1^\perp, \kappa_2^\perp <
\Lambda \end{array} \right.
\end{eqnarray}
Let us consider the quark distribution function $q(x) = {F_2(x) \over x}$
defined in Eq.(\ref{exact}). In presence of the ultraviolet cutoff
$\Lambda$, $q(x)$ depends on $\Lambda$ and schematically we have,
\begin{eqnarray}
q(x,\Lambda^2) = \sum \int_0^{\Lambda} \psi_2^2 + \sum \int_0^\Lambda
\int_0^\Lambda \psi_3^2.
\end{eqnarray}
For convenience, we write,
\begin{eqnarray}
q(x,\Lambda^2) =q_2(x,\Lambda^2) + q_3(x,\Lambda^2).
\end{eqnarray}
where the subscripts $2$ and $3$ denotes the two-particle and three-particle
contributions respectively.
Schematically we have,
\begin{eqnarray}
q(x,\Lambda^2) = && q(x,\mu^2) + \sum \int_\mu^\Lambda \mid \psi_2^h \mid^2
\nonumber \\
&&+ \sum \int_0^\mu \int_\mu^\Lambda \mid \psi_3^{sh} \mid^2 + 
\sum \int_\mu^\Lambda
\int_0^\mu \mid \psi_3^{hs} \mid^2 \nonumber \\
&& + \sum \int_\mu^\Lambda \int_\mu^\Lambda \mid \psi_3^{hh} \mid^2.
\end{eqnarray}
We investigate the contributions from the amplitudes $\psi_3^{sh}$ and
$\psi_3^{hs}$ to order $\alpha_s$ in the following.       
\subsection{Dressing with one gluon}

We substitute the Fock expansion Eq. (\ref{meson1}) in 
Eq. (\ref{rse}) and
make projection with a three particle state $b^\dagger (k_1,\sigma_1)
d^\dagger(k_2, \sigma_2) a^\dagger(k_3, \sigma_3) \mid 0 \rangle $ from the
left. In terms of the amplitudes $\psi_2$, $\psi_3$, we get,
\begin{eqnarray}
\psi_3^{\sigma_1 \sigma_2 \lambda_3}(x, \kappa_1; x_2, \kappa_2;
1-x-x_2, \kappa_3) = {\cal M}_1 + {\cal M}_2,
\end{eqnarray}
where 
\begin{eqnarray}
{\cal M}_1 = && { 1 \over E} (-) { g \over \sqrt{2 (2 \pi)^3}} T^a
{ 1 \over \sqrt{1 - x - x_2}} ~V_1~ 
\psi_2^{\sigma_1' \sigma_2}(1-x_2, -\kappa_2^\perp; x_2,\kappa_2^\perp), 
\end{eqnarray}
and
\begin{eqnarray}
{\cal M}_2 = && { 1 \over E} { g \over \sqrt{2 (2 \pi)^3}} T^a
{ 1 \over \sqrt{1 - x - x_2}} ~V_2~
\psi_2^{\sigma_1 \sigma_2'}(x,\kappa_1^\perp;1-x,-\kappa_1^\perp)
\end{eqnarray}
with
\begin{eqnarray}
E=
\big[ M^2  - {m^2 + (\kappa_1^\perp)^2 \over x} -
{m^2 + (\kappa_2^\perp)^2 \over x_2} - {(\kappa_3^\perp)^2 \over 1 - x -x_2}
\big ],
\end{eqnarray}       
\begin{eqnarray}
V_1=\chi_{\sigma_1}^\dagger \sum_{\sigma_1'}
\big [ { 2 \kappa_3^\perp \over 1 - x -x_2} - { (\sigma^\perp. \kappa_1^\perp
- i m) \over x} \sigma^\perp + \sigma^\perp {(\sigma^\perp. \kappa_2^\perp -
im) \over 1-x_2} \big] \chi_{\sigma_1'}. (\epsilon^\perp_{\lambda_1})^*,
\end{eqnarray}
and
\begin{eqnarray}
V_2=\chi_{-\sigma_2}^\dagger \sum_{\sigma_2'}
\big [ { 2 \kappa_3^\perp \over 1 - x -x_2} - \sigma^\perp
{ (\sigma^\perp. \kappa_2^\perp
- i m) \over y}  + \sigma^\perp {(\sigma^\perp. \kappa_1^\perp -
im) \over 1-x} \sigma^\perp 
\big] \chi_{-\sigma_2'}. (\epsilon^\perp_{\lambda_1})^*
\end{eqnarray}
\subsection{Perturbative analysis}
For $\kappa_1^\perp$ hard and $\kappa_2^\perp$ soft,
$\kappa_1^\perp+\kappa_2^\perp \approx \kappa_1^\perp$ and the multiple
transverse momentum integral 
over $\psi_3$ factorises into two independent integrals and the longitudinal
momentum fraction integrals become convolutions.
The contribution
from ${\cal M}_1$ to $\psi_3$ is,
\begin{eqnarray}
\psi_{3,1}^{\sigma_1,\sigma_2,\Lambda_3} (x,\kappa_1^\perp;x_2,\kappa_2^\perp;
1-x-x_2,-\kappa_2^\perp)= && - { g \over \sqrt{2 (2 \pi)^3}} T^a {x
\sqrt{1-x-x_2} \over 1 - x_2} {1 \over (\kappa_1^\perp)^2}
\nonumber \\ 
&& ~~ \chi^\dagger_{\sigma_1}\sum_{\sigma_{1}'}\big [ { 2 \kappa_1^\perp
\over 1-x-x_2}- {\sigma^\perp. \kappa_1^\perp \over x} \sigma^\perp
\big] \chi_{\sigma_1}'
. (\epsilon^\perp_{\lambda_1})^*
\nonumber \\
&&~~~~~\psi_2^{\sigma_1',\sigma_2}(1-x_2,-\kappa_2^\perp; x_2, \kappa_2^\perp).
\end{eqnarray}
Thus the contribution from ${\cal M}_1$ to the structure function is
\begin{eqnarray}
\sum \int \mid \psi_{3,1}^{hs} \mid^2 = {\alpha_s \over 2 \pi} C_f
ln{\Lambda^2 \over \mu^2} \int_x^1 {dy \over y} P_{qq}({x \over y})
q_2(y,\mu^2), 
\end{eqnarray}
where
\begin{eqnarray}
P_{qq}({x \over y}) = { 1 + ({x \over y})^2 \over 1 - {x \over y}}.
\end{eqnarray}

For the configuration $\kappa_1^\perp$ hard, $\kappa_2^\perp$ soft,
contribution from ${\cal M}_2$ does not factorize and the asymptotic
behavior of the integrand critically depends on the asymptotic behavior of
the two-particle wavefunction  $\psi_2$. To determine this behavior,
we have to analyze the bound state equation which shows that for large
transverse momentum $\psi_2 (\kappa^\perp) \approx { 1 \over
(\kappa^\perp)^2}$. Thus contribution from ${\cal M}_2$ for scale evolution
is suppressed by the bound state wavefunction. Analysis of the interference
terms (between ${\cal M}_1$ and $ {\cal M}_2$) shows that their
contribution also is suppressed by the bound state wavefunction.

For the configuration $\kappa_1^\perp$ soft, $\kappa_2^\perp$ hard,
contributions from ${\cal M}_1$ and the interference terms are suppressed by
the wave function. Contribution from ${\cal M}_2$ factorises both in
transverse and longitudinal space and generate a pure wavefunction
renormalization contribution:   
\begin{eqnarray}
\sum \int \mid \psi_{3,2}^{sh} \mid^2 = {\alpha_s \over 2 \pi} C_f 
ln{\Lambda^2 \over \mu^2} \int_0^1 dy {1 + y^2 \over 1-y}q_2(x,\mu^2).
\end{eqnarray} 

We have seen that even though the multi-parton contributions to the structure
function involve both coherent and incoherent phenomena, in the hard region
coherent effects are suppressed by the wavefunction.

\subsection{Corrections from normalization condition}
In the dressed quark calculation, we have seen that the singularity that
arises as $x
\rightarrow 1 $ from real gluon emission is canceled by the correction from
the normalization of the state (virtual gluon emission contribution
from wave function
renormalization). In the meson bound state calculation, so far we have
studied the effects of a hard real gluon emission. In this section we study
the corrections arising from the normalization condition of the quark
distribution in the composite bound state.

Collecting all the terms arising from the hard gluon emission contributing
to the quark
distribution function, we have,
\begin{eqnarray}
q(x,\Lambda^2) &&= q_2(x,\mu^2) + q_3(x,\mu^2) \nonumber \\
&& ~~~~~~ + { \alpha_s \over 2 \pi} C_f
ln{\Lambda^2 \over \mu^2} \int_x^1 {dy \over y} P_{qq} ({x \over y}) q_2(y,
\mu^2) \nonumber \\
&& ~~~~~~~ + {\alpha_s \over 2 \pi} C_f ln{\Lambda^2 \over \mu^2}
q_2(x,\mu^2) \int dy P(y).
\end{eqnarray}
We have a similar expression for the antiquark distribution function. 

The normalization condition on the quark distribution function should be
such that there is one valence quark in the bound state at any scale $Q$.
We choose the factorization scale $ \mu = Q_0$.
Let us first set the scale $ \Lambda = Q_0$. Then we have (in the truncated
Fock space)
\begin{eqnarray}
\int_0^1 dx q_2(x,Q_0^2) + \int_0^1 dx q_3(x ,Q_0^2) = 1.
\end{eqnarray}
Next set the scale $ \Lambda = Q$. We still require
\begin{eqnarray}
\int_0^1 dx q_2(x,Q^2) + \int_0^1 dx q_3(x,Q^2) = 1.
\end{eqnarray}
We note that the evolution of $q_3$ requires an extra hard gluon which is
not available in the truncated Fock space. Thus in the present approximation 
$ q_3(x,Q^2 ) = q_3(x,Q_0^2) $. 

Carrying out the integration explicitly, we arrive at 
\begin{eqnarray}
\int_0^1 dx q_2(x,\mu^2) \big[ 1 + {2 \alpha_s \over 2 \pi} C_f ln{Q^2 \over
Q_0^2} \int dy P(y)\big] + \int_0^1 dx q_3(x,Q^2) = 1
\end{eqnarray}

Thus we face the necessity to $``$renormalize" our quark distribution
function. Let us define a renormalized quark distribution function 
\begin{eqnarray}
q_2^R (x,Q_0^2) = q_2(x, Q_0^2) \big [1 + {\alpha_s \over 2 \pi} C_f ln{Q^2
\over Q_0^2} \int_0^1 dy P(y) \big]
\end{eqnarray}
so that, to order $\alpha_s$,
\begin{eqnarray}
\int_0^1 dx q_2^R(x,Q_0^2) + \int_0^1 dx q_3(x,Q_0^2) = 1.
\end{eqnarray}
We have,
\begin{eqnarray}
q_2(x,Q_0^2) = q_2^R(x,Q_0^2)  \big [ 1 - 2 {\alpha_s \over 2 \pi} C_f
ln{Q^2 \over Q_0^2} \int_0^1 dy P(y) \big]. 
\end{eqnarray}
Collecting all the terms, to order $\alpha_s$, we have the normalized quark
distribution function,
\begin{eqnarray}
q(x,Q^2)&& = q_2^R(x,Q_0^2) \nonumber \\
&& ~~~~~~+ {\alpha_s \over 2 \pi} C_f ln{Q^2 \over Q_0^2} \int_0^1 dy
q_2^R(y, Q_0^2) \int_0^1 dz \delta(zy-x) {\tilde P}(z) \nonumber \\
&& ~~~~~~ + q_3(x,Q^2) 
\end{eqnarray}
\
with $ {\tilde P}(z) = P(z) - \delta(z-1) \int_0^1 dy P(y)$.    
   
We see that just as in the dressed quark case, the singularity arising as $
x \rightarrow 1$ from real gluon emission is canceled in the quark
distribution function once the normalization
condition is properly taken in to account. From this derivation
we begin to recognize the lowest order term of the Altarelli-Parisi 
evolution equation.

\section{Discussion}
In this paper we have investigated the feasibility of analyzing deep
inelastic structure functions in the Hamiltonian formulation. 
Our starting point has been the 
non-perturbative
result obtained using equal-$x^+$ light-front current algebra and the
BJL limit which clearly shows exact Bjorken scaling if we forget the
non-trivial renormalization issues involved in 
the matrix element of the bilocal current
sandwiched between hadronic states. 
In this preliminary study we have focused on the leading contribution
(ignoring power suppressed terms) to the unpolarized structure function
$F_2$. Since the {\it good} (+) component of the bilocal current is
involved, 
expanding the hadronic state into
multi-parton wavefunctions, we obtained the structure function ${F_2 \over
x}$ as a sum of
the squared multi-parton wavefunction integrated over longitudinal and
transverse momenta. Taking into account the QCD interaction which 
connects the various wavefunctions and
scaling violation comes in the picture. We took the light-front
Hamiltonian and demonstrated how one could get logarithmic growth
 and the splitting
functions in the context of a dressed quark. The probabilistic
interpretation of the splitting function is very clear in this language. We
also noted how $x\rightarrow 0$ singularity could be taken care of by the
normalization of the state in this language.
In the bound state calculations involving a meson, we clearly illustrated
how factorization of the non-perturbative and perturbative parts in the
structure function occurred in our approach.
We have also illustrated how various coherent effects are power suppressed 
in the hard region which leads to the standard perturbative evolution of parton
distributions.

It is of interest to carry out higher order calculations for the dressed
quark to see explicitly how the running coupling constant comes into play in
this formulation. 

The ultimate test of the present formulation (success or failure)
is in its application  
to the
analysis of the so-called {\it higher-twist} structure functions. The
longitudinal structure function $F_L$ 
and the polarized structure function $G_T$ currently attracts lot of
attention. Even the questions of physical interpretation, 
factorization and evolution of these
structure functions are under intense investigation and 
debate at the present time. 
We plan to study these problems in the near future in  
the formulation presented in this paper which keeps close contact 
with physical intuitions of
the pre-QCD parton picture while maintaining the complexities of QCD.     
\acknowledgements
We acknowledge discussions with the members of the Theory Group, Saha
Institute of Nuclear Physics,
and the members of the Nuclear Theory Group, Physics Department, Iowa
State University. We also appreciate many useful conversations with Wei-Min
Zhang. This work was supported in part by the U.S. Department of
Energy under Grant No. DEFG02-87ER40371, Division of High Energy and Nuclear
Physics. 
\appendix
\section{Notations and Conventions}
We use the definitions
\begin{eqnarray}
x^\pm = x^0 \pm x^3, \, \, x^\perp = (x^1,x^2).
\end{eqnarray}
\begin{eqnarray}
\gamma^\pm =\gamma^0 \pm \gamma^3.
\end{eqnarray}.
\begin{eqnarray}
\psi^\pm = \Lambda^\pm \psi, \, {\rm with} \, \Lambda^\pm = { 1 \over 4}
\gamma^{\mp} \gamma^\pm.
\end{eqnarray}
The normalization of the state is
\begin{eqnarray}
\langle P \mid P' \rangle = 2 (2 \pi)^3 P^+ \delta^3 (P-P').
\end{eqnarray}
The volume elements 
\begin{eqnarray}
[dk] = {dk^+ d^2 k^\perp \over 2 (2 \pi)^3 \sqrt{k^+}}, 
\end{eqnarray}
\begin{eqnarray}
\{dk\} = {dk^+ d^2 k^\perp \over \sqrt{2 (2 \pi)^3 k^+}},
\end{eqnarray}
and
\begin{eqnarray}
(dk) = {dk^+ d^2 k^\perp \over 2 (2 \pi)^3 k^+}.
\end{eqnarray}  
\section{Normalization of state}
The normalization of the state $\mid P \rangle$ is 
\begin{eqnarray}
\langle P' \mid P\rangle = 2 (2 \pi)^3 P^+ \delta^3(P-P').
\end{eqnarray}
In the truncated space for the meson state, we have,
\begin{eqnarray}
\sum_{\sigma_1,\sigma_2}\int dk_1^+ d^2k_1^\perp \int dk_2^+ d^2k_2^\perp
\delta^3(P-k_1-k_2) \mid\phi_2(P \mid k_1,\sigma_1;k_2, \sigma_2) \mid^2
\nonumber \\
+\sum_{\sigma_1,\sigma_2,\lambda_3}\int dk_1^+ d^2k_1^\perp \int dk_2^+ d^2k_2^\perp
dk_3^+ d^2k_3^\perp 
\delta^3(P-k_1-k_2-k_3) 
\nonumber \\
~~~\times \mid \phi_3(P \mid k_1,\sigma_1;k_2, \sigma_2,
k_3, \lambda_3) \mid^2 = 1.
\end{eqnarray}
In terms of the amplitudes $ \psi_2, \psi_3$, the normalization condition in
the truncated Fock space sector reads
\begin{eqnarray}
\sum_{\sigma_1,\sigma_2} \int dx_1 d^2 \kappa_1^\perp \int dx_2 
d^2\kappa_2^\perp 
\delta(1-x_1-x_2) \delta^2(\kappa_1^\perp+ \kappa_2^\perp)
\mid \psi_2^{\sigma_1,\sigma_2} (x_1,\kappa_1^\perp; x_2, \kappa_2^\perp) 
\mid^2 +
\nonumber \\
\sum_{\sigma_1,\sigma_2, \lambda_3} 
\int dx_1 d^2 \kappa_1^\perp \int dx_2 
d^2\kappa_2^\perp  \int dx_3 d^2 \kappa_3^\perp
\delta(1-x_1-x_2-x_3) \delta^2(\kappa_1^\perp+ \kappa_2^\perp +
\kappa_3^\perp)
\nonumber \\
~~~ \times \mid \psi_3^{\sigma_1,\sigma_2, \lambda_3} 
(x_1,\kappa_1^\perp; x_2, \kappa_2^\perp; x_3,\kappa_3^\perp) 
\mid^2=1. \label{norm}
\end{eqnarray} 
 
\section{Operator definition of gluon
distribution function and the longitudinal momentum sum rule}

Consider the conserved  longitudinal momentum operator 
\begin{eqnarray}
P^+ = { 1 \over 2} \int dx^- d^2 x^\perp \theta^{++}
\end{eqnarray}
where
\begin{eqnarray}
\theta^{++} = \theta^{++}_F + \theta^{++}_G .
\end{eqnarray}
The fermion contribution 
\begin{eqnarray}
\theta^{++}_F = i {\bar \psi} \gamma^+ \partial^+ \psi 
\end{eqnarray}
and the gluon contribution
\begin{eqnarray}
\theta^{++}_G = F^{+ \nu a} F_{\nu}^{+a}
\end{eqnarray}
where
\begin{eqnarray}
F^{\mu \nu a} = \partial^{\mu} A^{\nu a} - \partial^{\nu} A^{\mu a} + g
f^{abc} A^{\mu b} A^{\nu c}.
\end{eqnarray}

From the definition
\begin{eqnarray}
F_2^F(x) = {x \over 4 \pi P^+} \int d \eta e^{-i \eta x} 
\langle P \mid \big [{\bar \psi}(y) \gamma^+ \psi(0) - {\bar \psi}(0)
\gamma^+ \psi(y) \big ] \mid P \rangle, 
\end{eqnarray}
with $ \eta ={ 1 \over 2} P^+ y^-$, we have
\begin{eqnarray}
\int dx F_2^F(x) = ({1 \over P^+})^2 \langle P \mid \theta^{++}_F \mid P
\rangle .
\end{eqnarray}
Formally, we can define the $``$gluon structure function" \cite{CSS} 
\begin{eqnarray}
F_2^G(x) = { 1 \over 4 \pi P^+} \int dy^- e^{- {i \over 2} P^+ y^- x } 
\langle P \mid F^{+ \nu a} (y^-)F^{+a}_{\nu}(0) \mid P \rangle .
\end{eqnarray}
We have,
\begin{eqnarray}
\int dx F_2^G(x) = ({ 1 \over P^+})^2 \langle P \mid \theta^{++}_G \mid P
\rangle 
\end{eqnarray}
and the momentum sum rule
\begin{eqnarray}
\int dx \big [ F_2^F + F_2^G \big ] =1.
\end{eqnarray}

Next we explicitly verify the longitudinal momentum sum rule in the
truncated Fock space.
The fermionic part of the longitudinal momentum operator
\begin{eqnarray}
P_F^+ = { 1 \over 2} \int dx^- d^2 x^\perp 2 i (\psi^+)^\dagger \partial^+
\psi^+ .
\end{eqnarray}
Using the field expansion
\begin{eqnarray}
\psi^+(x) = \sum_{\lambda} \chi_{\lambda} \int 
[dp]
 \big [ b(p,\lambda) e^{-i
p.x} + d^{\dagger}(p,-\lambda) e^{ip.x} \big ],
\end{eqnarray}
\begin{eqnarray}
P^+_q = \sum_{\lambda} \int (dp) p^+\big [ b^\dagger(p,\lambda)
b(p,\lambda) + d^\dagger(p,\lambda) d (p,\lambda) \big] ,
\end{eqnarray}
with $ (dp)$ defined in appendix A.

In $A^{+a}=0$ gauge, the gluonic part of the longitudinal momentum operator 
\begin{eqnarray}
P^{+}_G = { 1 \over 2} \int dx^- d^2 x^\perp \partial^+ A^{ja} \partial^+
A^{j a}.
\end{eqnarray}
Using the field expansion 
\begin{eqnarray}
A^{ja} = \sum_{\lambda} \int (dq) \big[ \epsilon^j_{\lambda} a^a(q, \lambda)
e^{-i q.x} + (\epsilon^j_\lambda)^* (a^a)^\dagger(q,\lambda) e^{iq.x} \big],
\end{eqnarray}
\begin{eqnarray}
P_G^+ = \sum_{\lambda} \int (dq) q^+(a^a)^\dagger(q,\lambda) a^a(q,\lambda).
\end{eqnarray}

We have the relation
\begin{eqnarray}
\langle P' \mid P^+ \mid P \rangle = \big [ 2 (2 \pi)^3 P^+ \delta^3(P - P')
\big ] P^+.
\end{eqnarray}
Explicit evaluation in the truncated Fock space gives

\begin{eqnarray}
\sum_{\sigma_1,\sigma_2} \int dx_1 d^2 \kappa_1^\perp \int dx_2 
d^2\kappa_2^\perp 
\delta(1-x_1-x_2) \delta^2(\kappa_1^\perp+ \kappa_2^\perp)
\nonumber \\
~~~~ \times (x_1+x_2)\mid \psi_2^{\sigma_1,\sigma_2} (x_1,\kappa_1^\perp; x_2, \kappa_2^\perp) 
\mid^2 +
\nonumber \\
\sum_{\sigma_1,\sigma_2, \lambda_3} 
\int dx_1 d^2 \kappa_1^\perp \int dx_2 
d^2\kappa_2^\perp  \int dx_3 d^2 \kappa_3^\perp
\delta(1-x_1-x_2-x_3) \delta^2(\kappa_1^\perp+ \kappa_2^\perp +
\kappa_3^\perp) 
\nonumber \\ 
~~~~ \times (x_1+x_2+x_3)
\mid \psi_3^{\sigma_1,\sigma_2, \lambda_3} 
(x_1,\kappa_1^\perp; x_2, \kappa_2^\perp; x_3,\kappa_3^\perp) 
\mid^2=1, 
\end{eqnarray} 
which is automatically satisfied because of the normalization condition
given in Eq.(\ref{norm}). 
Thus momentum fraction sum rule is trivially satisfied given the
normalization condition on the amplitudes $ \psi_2, \psi_3$ etc.

\end{document}